Hyunjun Ma, Jin-soo Kim, Jong-Ho Choe, and Q-Han Park

# Deep-learning-assisted reconfigurable metasurface antenna for real-time holographic beam steering

**Abstract:** We propose a metasurface antenna capable of real-time holographic beam steering. An array of reconfigurable dipoles can generate on-demand far-field patterns of radiation through the specific encoding of meta-atomic states i.e., the configuration of each dipole. Suitable states for the generation of the desired patterns can be identified using iteration, but this is very slow and needs to be done for each far-field pattern. Here, we present a deep-learning-based method for the control of a metasurface antenna with point dipole elements that vary in their state using dipole polarizability. Instead of iteration, we adopt a deep learning algorithm that combines an autoencoder with an electromagnetic scattering equation to determine the states required for a target far-field pattern in real-time. The scattering equation from Born approximation is used as the decoder in training the neural network, and analytic Green's function calculation is used to check the validity of Born approximation. Our learning-based algorithm requires a computing time of within 200 μs to determine the meta-atomic states, thus enabling the real-time operation of a holographic antenna.

## 1. Introduction

Beam steering without requiring the mechanical movement of antenna elements is an important component of electromagnetic (EM) wave applications. To achieve this, a phased array controls the phase of each antenna element and shapes the wavefront to steer the beam direction, which leads to high power consumption and requires complex electronics. Recently, reconfigurable metasurfaces have received interest as promising candidates for low-cost beam-steering antennas [1-9]. As a two-dimensional array of unit elements with variable states, a reconfigurable metasurface can modulate the wavefront via the collective scattering of elements and achieve beam steering by varying the state of the unit elements. On-demand beam steering requires an algorithm that can determine the corresponding state of the unit elements based on a far-field map. For general holographic far-field maps, only numerical solutions are permitted [10-11], which are usually found using iterative optimization methods such as genetic algorithms (GAs) [12-13] or particle swarm methods [14] and the Gerchberg-Saxton(GS) algorithm [15]. However, iterative methods take a long time to arrive at the solution and must be performed for each far-field map. Deep learning is a potent replacement method for reconfigurable metasurfaces [16-22] that has been used for the creation of holograms with experimental verification [17], the beamforming of the antenna [18-19], and the adaptive invisible cloak [22].

In the present study, we present a deep-learning-based algorithm for real-time holographic beam steering using a reconfigurable metasurface. We adopt an autoencoder neural network [17, 23-24] in which the encoder generates the element states from a far-field map, while the physics-assisted decoder directly solves the scattering equation to obtain a far-field map from the generated states instead of undergoing neural network training. For simplicity, we assume the unit elements to be point dipoles and control their reconfigurable states by varying their polarizability. To ensure the stable training of the neural network, we employ higher-order Born approximation [25-26] of the scattering equation and test its accuracy by comparing it with analytic Green's function calculations [27-31]. We investigated the effect of multiple scattering and its dependence on the dipole polarizability, as well as the quality of the far-field pattern generated, by varying the Born approximation order up to 3. Our autoencoder model is trained using handwritten digit data from the Modified National Institute of Standards and Technology (MNIST) database [32] in order to minimize the difference between the input MNIST image and the decoder-generated far-field map. We test our model with various far-field maps, including multi-directional beams and holographic maps. Our neural network generates meta-unit states that can produce high-fidelity holographic far-field patterns. Additionally, our trained neural network was able to generate untrained target images, such as a focused multi-beam, indicating that it acts as an inverse operation of the electromagnetic scattering equation.

For a metasurface with 900 unit elements, our model determines the required states within 200 μs. As a result, our learning-based algorithm enables the real-time operation of a reconfigurable metasurface for a holographic antenna.

## 2. Reconfigurable metasurface
### a. Dipole approximation and electromagnetic scattering

We model a reconfigurable metasurface as a two-dimensional array of dipoles with complex polarizability $\alpha_n$ for the $n$-th dipole. In reconfiguring the unit elements, we can vary both the amplitude and phase $\alpha_n$ Here, for simplicity, we fix the phase at either 0 or $\pi$ and vary the amplitude to a maximum of $\alpha_{\max}$. In other words, we consider only the real $\alpha_n = k_n \alpha_{\max}, -1 \leq k_n \leq 1$. Later, we will also consider the binary case where $\alpha_n = \pm \alpha_{\max}$. Our metasurface consists of $30 \times 30$ dipoles rectangularly arranged and spaced 0.2 wavelengths apart in each direction. We assume that the $n$-th dipole ($n = 1, 2, 3 \ldots N = 900$) with polarizability $\alpha_n$ located at $\mathbf{r}_n$ had the induced polarization $\mathbf{P}_n = \alpha_n \mathbf{E}(\mathbf{r}_n)$. The total field $\mathbf{E}(\mathbf{r}_i)$ at $\mathbf{r}_i$ is the sum of the incident and dipole-scattered fields: [21]

$$\mathbf{E}(\mathbf{r}_i) = \mathbf{E}_{\text{inc}}(\mathbf{r}_i) + \sum_n \overleftrightarrow{\mathbf{G}}(\mathbf{r}_i, \mathbf{r}_n) \mathbf{P}_n \tag{1}$$

where $\overleftrightarrow{\mathbf{G}}(\mathbf{r}_i, \mathbf{r}_j)$ denotes the dyadic Green's function of the free space. In theory, combining the equation (1) and $\mathbf{E}(\mathbf{r}_n) = \alpha_n^{-1} \mathbf{P}_n$, the induced polarization for each dipole can be directly calculated from the matrix equation

$$\left[ \alpha_j^{-1} \delta_{jk} - \overleftrightarrow{\mathbf{G}}(\mathbf{r}_j, \mathbf{r}_k) \right] \mathbf{P}_k = \mathbf{E}_{\text{inc}}(\mathbf{r}_j), \tag{2}$$

where $\delta_{jk}$ is a Kronecker delta. However, in practice, the inverse process often becomes unstable when training the neural network and thus should be avoided. Instead, we use a recursive Born approximation for the training of the neural network and later check its accuracy by comparing its results for the generated dipole states with those derived from Green's function. Born approximation up to the third order is used to determine the far-field map for a given dipole polarizability set $\alpha_n, n = 1, \ldots, N$, [25]

$$\begin{aligned} \mathbf{E}(\mathbf{r}_i) = & \mathbf{E}_{\text{inc}}(\mathbf{r}_i) + \sum_j \overleftrightarrow{\mathbf{G}}(\mathbf{r}_i, \mathbf{r}_j) \alpha_j \mathbf{E}_{\text{inc}}(\mathbf{r}_j) \\ & + \sum_j \sum_k \overleftrightarrow{\mathbf{G}}(\mathbf{r}_i, \mathbf{r}_j) \alpha_j \overleftrightarrow{\mathbf{G}}(\mathbf{r}_j, \mathbf{r}_k) \alpha_k \mathbf{E}_{\text{inc}}(\mathbf{r}_k) \\ & + \sum_j \sum_k \sum_l \overleftrightarrow{\mathbf{G}}(\mathbf{r}_i, \mathbf{r}_j) \alpha_j \overleftrightarrow{\mathbf{G}}(\mathbf{r}_j, \mathbf{r}_k) \alpha_k \overleftrightarrow{\mathbf{G}}(\mathbf{r}_k, \mathbf{r}_l) \alpha_l \mathbf{E}_{\text{inc}}(\mathbf{r}_l) \end{aligned} \tag{3}$$

Figure 1 presents a schematic diagram of our metasurface antenna. The numerical values of $\alpha_n$ are chosen in the range $10^{-7}$ m$^3$ < $|\alpha_n/\epsilon_0| \leq 10^{-5}$ m$^3$ which is valid for our desired frequency range and the unit size [33-37], but it should be mentioned that our approach is universal and can be used in various systems with different scales. In training the neural network, an incident wave is generated by a 2.6 GHz dipole feed antenna located one wavelength away from the center of the metasurface, which is taken as the origin of the coordinates. This fixes the electric field $\mathbf{E}_{\text{inc}}(\mathbf{r}_j)$ and Green's function $\overleftrightarrow{\mathbf{G}}(\mathbf{r}_i, \mathbf{r}_j)$ except for the dipole vector of the feed antenna. We assume that the feed antenna is designed not to interfere with the scattered fields at the detection position so that the far-field map is generated by the scattered fields. We train the neural network to determine the states of the dipoles $[k_1, k_2, ..., k_N]$ that generate a far-field map that agrees with the input map.

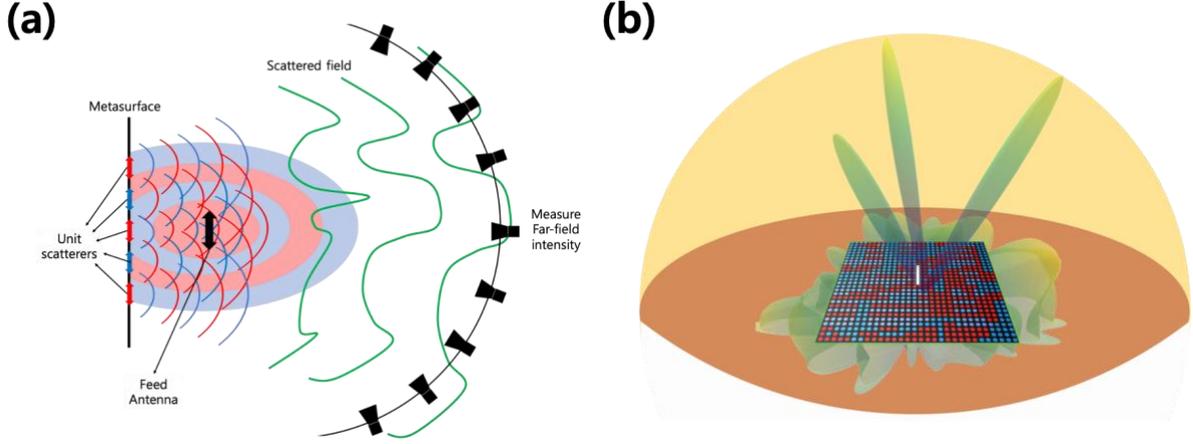

**Fig. 1**. (a) Schematic of the proposed metasurface antenna. The red and blue double arrows represent unit dipoles with a state of 1 and –1, respectively. The metasurface unit scatterers reflect and interfere with the source wave. We measure the far-field intensity for a (b) 30 x 30 metasurface antenna array. The metasurface modulates the dipole antenna source beam and generates an on-demand far-field map.

### b. Neural network architecture

We adopt a deep autoencoder neural network consisting of an encoder and a decoder. The encoder generates the dipole states from an input holographic far-field map and the decoder calculates the far-field map from the generated dipole states. The autoencoder is trained to minimize the error between the input and calculated far-field maps. During training, the encoder generates the optimal dipole state $[k_1, k_2, ..., k_N]$, which in turn generates the on-demand far-field.

Figure 2 shows a schematic diagram of the network architecture. We consider each input image to be a $u$–$v$ far-field map, which is a projection of the spherical surface map onto the x-y plane. We use MNIST handwritten digit images and resize them to a pixel size of (64, 64). The $u$–$v$ far-field map is only defined in the unit circle $u^2 + v^2 \leq 1$, and we set the pixel values outside this unit circle to be 0. We normalize the sum of the image pixel values to 1 because we are concerned with the directivity of the antenna.

We employ the Residual Network (ResNet) [38] architecture as our encoder model. The encoder consists of one input convolution layer, three ResNet blocks, and one fully connected layer. Each ResNet block consists of three convolution layers with 64 channels. We use the leakyReLU [39] function as the activation function for the convolution layers. The output dimension of the fully connected layer is (1, 900), which is the same as the number of metasurface unit elements and thus the number of dipoles. For the encoder output to represent dipole state $k_n$, we use the hyperbolic tangent activation function after a fully connected layer so that the value of the output vector component remains between –1 and 1.

We do not apply the neural network architecture to the decoder but rather solve the forward scattering equation (4) with the dipole polarizability $\alpha_n$ given by state $k_n$. We place a set of detectors sufficiently far away from the origin and evaluate the intensity of the scattered electric field $\mathbf{E}_{\text{sca}}(\mathbf{r}_{ij})$. The position of each detector is given by $\mathbf{r}_{ij} = R\, u_i \hat{x} + R\, v_j \hat{y} + R\sqrt{1 - u_i^2 - v_j^2}\hat{z}$ for $R = 10^5$ m, which is defined in the region $u_i^2 + v_j^2 \leq 1$ with $u_i = -1 + 2i/64$, $v_j = -1 + 2j/64$. We use the modified mean squared error (MSE) loss function for the normalized intensity of the scattered fields given by

$$L = \sum_{i=1}^{64} \sum_{j=1}^{64} \left( \frac{|\mathbf{E}_{\text{sca}}(\mathbf{r}_{ij})|^2}{\sum_{l=1}^{64} \sum_{m=1}^{64} |\mathbf{E}_{\text{sca}}(\mathbf{r}_{lm})|^2} - \hat{z}_{ij} \right)^2 \tag{5}$$

where $\hat{z}_{ij}$ is the pixel value for the input image at index $(i,j)$. Because loss function $L$ contains the absolute square of the complex-valued scattered fields and is thus not holomorphic in $\mathbf{E}$, care is needed in the backpropagation process. Instead of dealing with the derivative with respect to the complex quantity, we take the real and the imaginary parts separately as independent quantities and also take the derivative separately. For example, the derivative of the loss function with regard to state variable $k_n$ is

$$\frac{\partial L}{\partial k_n} = \sum_{ij} \frac{\partial L}{\partial \text{Re}\left[\mathbf{E}_{\text{sca}}(\mathbf{r}_{ij})\right]} \frac{\partial \text{Re}\left[\mathbf{E}_{\text{sca}}(\mathbf{r}_{ij})\right]}{\partial \alpha_n} \frac{\partial \alpha_n}{\partial k_n} + \frac{\partial L}{\partial \text{Im}\left[\mathbf{E}_{\text{sca}}(\mathbf{r}_{ij})\right]} \frac{\partial \text{Im}\left[\mathbf{E}_{\text{sca}}(\mathbf{r}_{ij})\right]}{\partial \alpha_n} \frac{\partial \alpha_n}{\partial k_n} \quad (6)$$

We use the Julia programming language and the Zygote package to calculate the derivative of the complex numbers by considering them as a pair of real numbers. We trained our proposed neural network structure and compared its meta-unit solutions with those obtained from other optimization techniques such as GA, and modified GS algorithms of [15]. We used the MNIST dataset, splitting it into training, validation, and test sets, each comprising 50,000, 10,000, and 10,000 samples, respectively. The third-order Born approximation was used for training with the Adam [40] optimizer having an initial learning rate of $1.0 \times 10^{-4}$ and a batch size of 128. We also used early stopping, and the training converged after 30 epochs, taking about 35 minutes with an Nvidia A6000 GPU. For GA optimization, we used a population size of 200, elitism rates of 0.2, and mutation rates of 0.1. We performed calculations using GA or the GS algorithms with scattering equation (3), using an Nvidia A6000 GPU. In the GS algorithm, we selected the meta-unit states with the lowest MSE loss after iterating the process until one of the already appeared binary states reappeared.

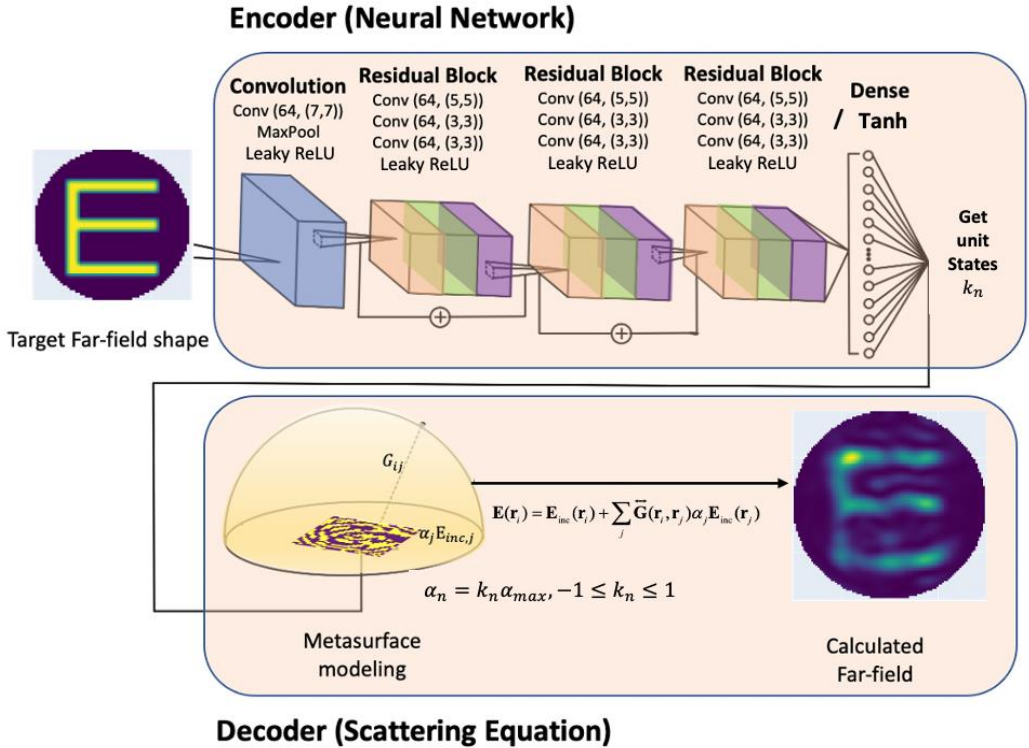

**Fig. 2**. Schematic diagram of the architecture of our autoencoder neural network. The encoder generates the meta-units states, and the decoder reproduces the on-demand far-field. We use the ResNet structure for our encoder and a scattering equation with Born approximation for our decoder.

## 3. Results and Discussion

Figure 3 presents four test examples using the scattering equation with third-order Born approximation and maximum polarizability of $\alpha_{max}/\epsilon_0 = 1.0 \times 10^{-6}$ m$^3$. Although our autoencoder model is trained using MNIST digit data only, we test the model using non-digit-type far-field maps and single- and multi-directional beaming and holographic images (Figure 3a). Figure 3b displays the generated dipole states and the far-field maps predicted by the trained autoencoder. In comparison with the MNIST images, the single-beam, multi-beam, and letter "E" images exhibit equally good recovery, suggesting that the encoder is not overfitted to MNIST-type far-field maps but has been trained for more general inverse operations. The resolution of the far-field maps is constrained by diffraction since we utilized a metasurface with a small (6 wavelengths × 6 wavelengths) size. Nonetheless, Figure 3b demonstrates good recovery.

In real applications, reconfiguring a metasurface by continuously varying the states is difficult to achieve. Thus, we also consider the binary truncation of the states generated by the model by taking the sign function of the states and evaluating the resulting far-field map (Figure 4). Despite the truncation, the predicted far-field maps are in reasonable agreement with the original input images. Though the holographic "E" image has side-lobe errors due to the diffraction caused by truncation, the binary approximation of our model outperforms the iterative GA in terms of directional beam steering. Figure 4b presents the binary states and far-field maps generated by the GA. The GA searches the optimal binary states by minimizing the MSE loss in (5). For a directional beam steering focusing on multiple spots, the GA easily falls into local minima missing certain spots when only the MSE loss function is used without additional regularizations. Figure 4c describes the results of the modified GS algorithm. Note the distributional similarity of the metasurface unit pattern between ours and the GS algorithm. Further, our neural network spends an average of 185 μs to generate 900 meta-units, while the GA requires an average of 232 sec, and the GS algorithm requires an average of 1.25 sec.

We also checked the validity of the Born approximation by increasing the iteration order and the maximum polarizability $\alpha_{max}$ (Figure 5). For $\alpha_{max}/\epsilon_0 = 1.0 \times 10^{-7}$ m$^3$ case, the effect of multiple scattering by nearby dipoles is negligible and thus the neural network can be trained with only first-order Born approximation. However, for a larger $\alpha_{max}/\epsilon_0 = 1.0 \times 10^{-5}$ m$^3$, first-order Born approximation generates a large side-lobe (Figure 5c), which tends to disappear as the Born approximation order increases. This is also reflected in the reduction in the MSE loss during training.

While our proposed algorithm is currently limited to a theoretical prototype, it can be applied to real-world metasurface antennas. To implement our algorithm, one would need to extract the polarizability of the metasurface unit [33-37], which can be done through various methods such as calculating the far field scattered from the designed metasurface unit in experiments or numerical simulations and optimizing the polarizability to generate that field. Once we have the polarizabilities for all possible states, we can establish the scattering equation for the antenna structure. By training the neural network encoder to be an inverse operation of the scattering equation, we can obtain the proper solution for the desired far field to be generated. Therefore, although our current focus is on the theoretical prototype, our proposed algorithm has practical applications in the design and optimization of metasurface antennas.

## 4. Conclusion

We proposed a deep-learning-based method for the control of a reconfigurable metasurface antenna. We modeled the metasurface as a collection of dipoles with states of varying polarizability and used a deep autoencoder neural network combined with a scattering equation and Born approximation to generate on-demand far-field maps. Our proposed autoencoder exhibited high accuracy and a much faster speed compared with the conventional GA approach and the GS algorithm. This would allow for the real-time operation of a reconfigurable metasurface antenna for beam steering.

Because our model simplifies the reconfigurable metasurface elements as dipoles with varying states, the realistic application of our model requires further consideration. In a real device, the finite size effect of unit elements should be considered, which goes beyond dipole approximation. We employed dipole polarizability with varying amplitude and the phase fixed to 0 or $\pi$, but this was not essential. We could have kept the amplitude fixed and varied the phase or varied both. When building a device element that represents a dipole with a variable state, it is preferable to employ a discrete state, such as a binary one. We demonstrated that the binary truncation of our continuous autoencoder model still produced a reasonable performance. The accuracy of our approach could be further improved if we employed a discrete state in our autoencoder model from the beginning, without truncation afterward. This can be achieved by modifying the normalization procedure, which will be considered in future research. Our work can easily be extended to more general metasurface antennas, phased arrays, and other far-field imaging applications.

## 5. Acknowledgement

This work was supported by the National Research Foundation of Korea (NRF) grant funded by the Korea government (MSIT) (No. 2021R1A2C2008814) and the Global Frontier Program. (CAMM-2014M3A6B3063710)

# Figures and table

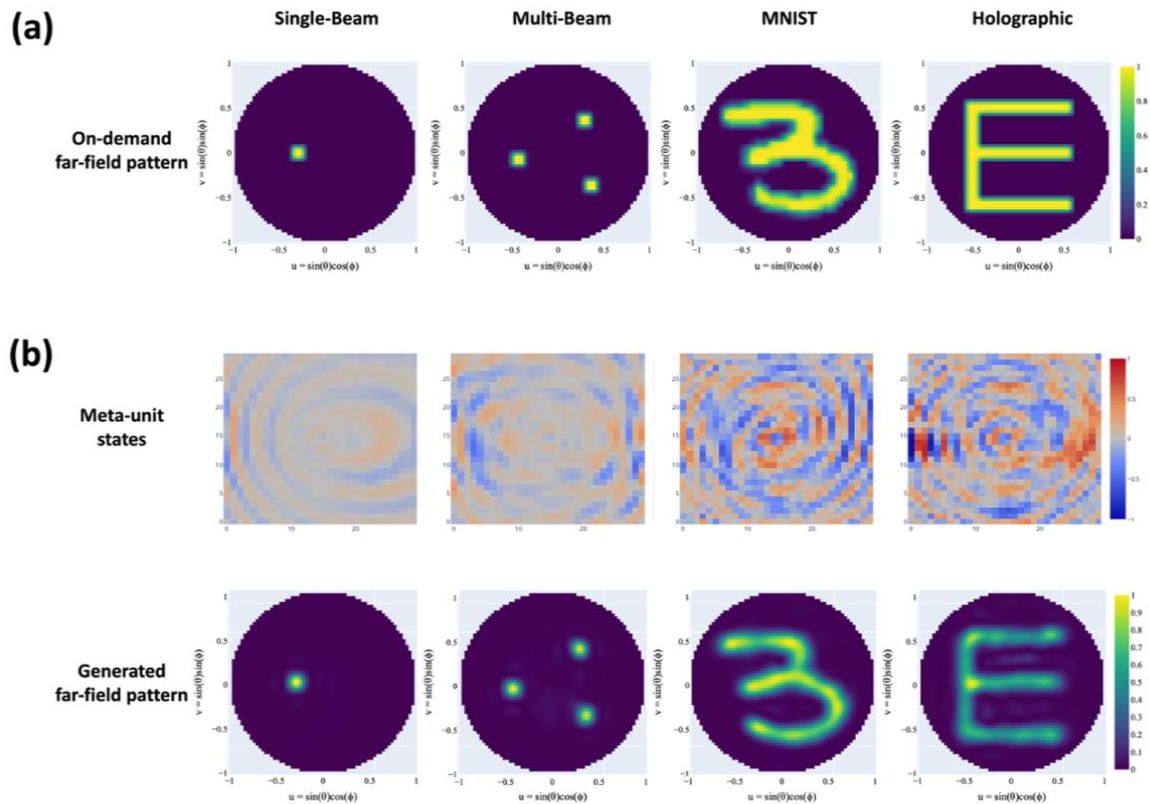

**Fig 3.** Target far-field map, the state pattern generated by the neural network, and the calculated far-field map from the analytic Green's function (a) On-demand far-field maps: single-beam, multi-beam, MNIST, and "E" images. (b) Meta-unit states and the generate far-field pattern if the meta-unit has a continuous state.

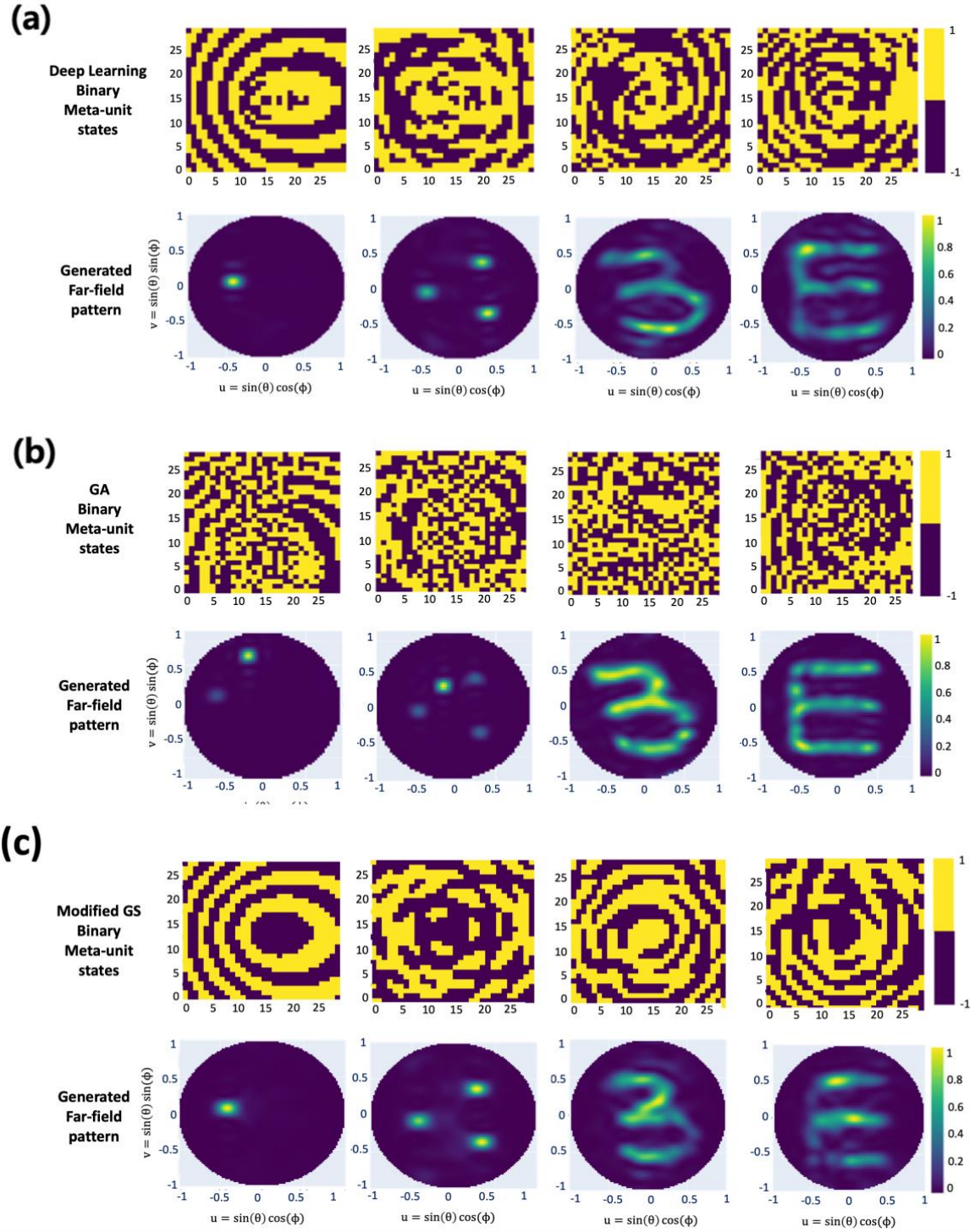

**Fig. 4.** Comparison of the results from the proposed neural network and the genetic algorithm when trained with the MSE loss from the far-field maps. (a) Binary meta-unit states of –1 or 1 generated by the proposed neural network. (b) Meta-unit states and the far-field pattern generated by the genetic algorithm. (c) Meta-unit states and the far-field pattern generated by the GS algorithm.

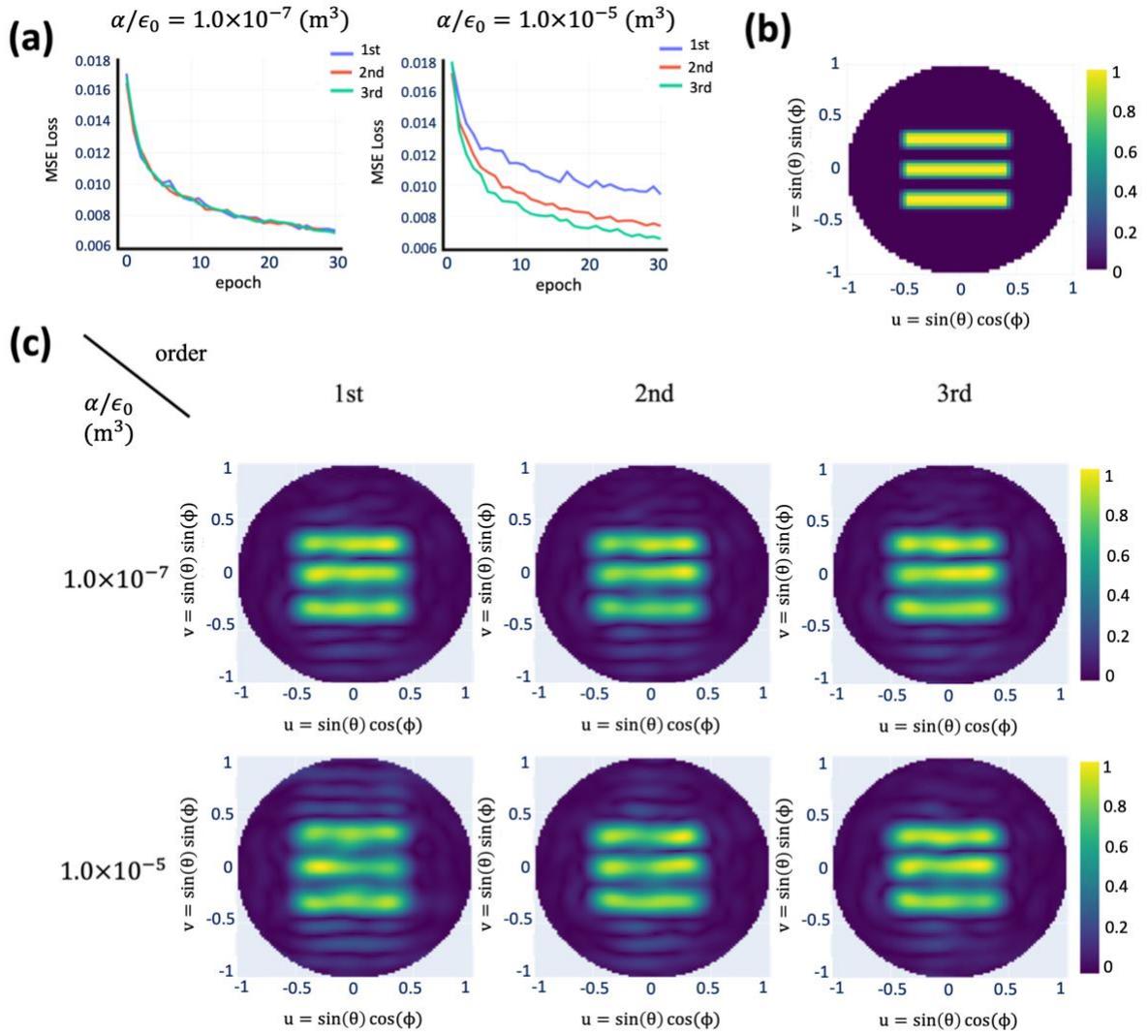

**Fig. 5.** Loss function of the trained autoencoder and the far-field calculated using the analytic Green's function method according to the magnitude of the polarizability and order of Born approximation used in the decoder. (a) Loss function of the neural network. (b) Target far-field map. (c) Generated far-field map. Note that for $\alpha_{max}/\epsilon_0 = 1.0 \times 10^{-5}$ (m$^3$), first-order Born approximation generated strong side lobes compared to the second and third order approximation.

| Methods | Calculation time | |
|---|---|---|
| | Training time | Calculation time |
| **AutoEncoder & first-order Born** | 1854 sec | 194 μs |
| **AutoEncoder & second-order Born** | 1921 sec | 192 μs |
| **AutoEncoder & third-order Born** | 2077 sec | 185 μs |
| **Genetic Algorithm** | 232 sec | |
| **Gerchberg-Saxton** | 1.25 sec | |

**Table 1.** Calculation time according to the proposed methods, calculation time except Genetic Algorithm is averaged over the MNIST dataset. Genetic algorithm calculation time is averaged over the cases of Fig. 4.